\documentclass[12pt]{article}
\usepackage{amssymb}
\usepackage{amsmath}
\usepackage{apacite}
\usepackage{latexsym}
\usepackage{epsfig}
\usepackage{graphicx}
\usepackage{subfigure}
\usepackage{wasysym}
\usepackage{threeparttable}
\usepackage{natbib}
\usepackage{color, soul}
\usepackage{epstopdf}
\usepackage{bm}
\usepackage{float}
\usepackage{todonotes}
\usepackage{verbatim}
\usepackage{booktabs}

\usepackage{hyperref}

\def\boxit#1{\vbox{\hrule\hbox{\vrule\kern6pt
          \vbox{\kern6pt#1\kern6pt}\kern6pt\vrule}\hrule}}

\def\bse{\begin{eqnarray*}}
\def\ese{\end{eqnarray*}}
\def\be{\begin{eqnarray}}
\def\ee{\end{eqnarray}}
\def\bq{\begin{equation}}
\def\eq{\end{equation}}
\def\bse{\begin{eqnarray*}}
\def\ese{\end{eqnarray*}}

\begin{document}

\thispagestyle{empty} 
\baselineskip=28pt

\begin{center}
{\LARGE{\bf The Link Between Health Insurance Coverage and Citizenship Among Immigrants: Bayesian Unit-Level Regression Modeling of Categorical Survey Data Observed with Measurement Error}}

\end{center}

\baselineskip=12pt

\vskip 2mm
\begin{center}

Paul A. Parker\footnote{(\baselineskip=10pt to whom correspondence should be
  addressed) Department of Statistics, University of California, Santa Cruz, 1156 High Street,
  Santa Cruz, CA 95064, paulparker@ucsc.edu}\ , 
Scott H. Holan\footnote{\baselineskip=10pt Department of Statistics, University of
  Missouri, 146 Middlebush Hall, Columbia, MO 65211,
  holans@missouri.edu}\,\footnote{\baselineskip=10pt Office of the Associate Director for Research and Methodology, U.S. Census Bureau, 4600 Silver
  Hill Road, Washington, D.C. 20233, scott.holan@census.gov}\ ,
  James D. Bachmeier\footnote{\baselineskip=10pt Department of Sociology, Temple University, 713 Gladfelter Hall,
Philadelphia, PA 19122, james.bachmeier@temple.edu}\ , and
  Claire Altman\footnote{\baselineskip=10pt Department of Health Sciences, University of Missouri, 322 Clark Hall, Columbia, MO 65211, \\ altmanc@health.missouri.edu}

\end{center}

\vskip 4mm
\baselineskip=12pt 
\begin{center}
{\bf Abstract}
\end{center}

Social scientists are interested in studying the impact that citizenship status has on health insurance coverage among immigrants in the United States. This can be done using data from the Survey of Income and Program Participation (SIPP); however, two primary challenges emerge. First, statistical models must account for the survey design in some fashion to reduce the risk of bias due to informative sampling. Second, it has been observed that survey respondents misreport citizenship status  at nontrivial rates. This too can induce bias within a statistical model. Thus, we propose the use of a weighted pseudo-likelihood mixture of categorical distributions, where the mixture component is determined by the latent true response variable, in order to model the misreported data. We illustrate through an empirical simulation study that this approach can mitigate the two sources of bias attributable to the sample design and misreporting. Importantly, our misreporting model can be further used as a component in a deeper hierarchical model. With this in mind, we conduct an analysis of the relationship between health insurance coverage and citizenship status using data from the SIPP.

\baselineskip=12pt
\par\vfill\noindent
\par\medskip\noindent
\clearpage\pagebreak\newpage \pagenumbering{arabic}
\baselineskip=24pt

\section{Introduction}

Social scientists specializing in the foreign-born population of the United States have long focused on questions related to citizenship that have broad academic and policy significance. Such research seeks to understand the trends and dynamics of naturalization itself \citep{amuedo2021recent, le2022family, ryo2022importance, yang1994explaining} and increasingly, how the acquisition of U.S. citizenship through naturalization is linked to other key indicators of economic and social integration and material well-being \citep{altman2021material, borjas2003welfare, buchmueller2007immigrants, hainmueller2015naturalization}.

Social scientists hypothesize that foreign-born survey respondents misreport their citizenship and/or legal status at a substantial rate. That is, foreign-born respondents either intentionally or unknowingly inaccurately report their status. Relatively large biases stemming from reporting error lend to uncertainty in estimates of the size and characteristics of the naturalized citizen population derived from survey data \citep{brown2019predicting, van2013well}.  Moreover, reporting error of citizenship can also bias estimates of the association between naturalization and indicators of integration and well-being such as health insurance coverage or poverty. The degree of bias introduced is a question that is of increasing salience in the social scientific literature on immigration and immigrant integration \citep{donato2011we,young2017documenting}.

Despite tacitly acknowledging misreporting, scholars studying the foreign-born population often implicitly assume that there is no bias attributable to measurement error when studying immigration-related measures of citizenship status by analyzing public-use national data sources such as the Survey of Income and Program Participation (SIPP) and American Community Survey (ACS) released by the U.S. Census Bureau. To date, the limited empirical evidence suggests that status measures are prone to substantial reporting error, due either to misrepresentation or confusion over question meanings \citep{brown2019predicting, van2013well}. 

For example, using the 1990 census and the 1996 Current Population Survey (CPS), \cite{passel1998immigrants} found that survey respondents commonly misreported their citizenship status and frequently reported being citizens, as evidenced by their insufficient duration of U.S. residence required to naturalize (see also, \cite{brown2019predicting}). This occurred primarily among foreign-born respondents who had resided in the U.S. for less than five years, but reported being naturalized, however this was not uncommon among longer term immigrants as well. 

Furthermore, \cite{van2013well} used the 2010 one year ACS public use microdata sample and compared national rates of naturalization to those estimated using data from the Office of Immigration Statistics and found higher reported rates in the ACS data for both short (less than 5 years) and longer-term foreign-born respondents. \citet{brown2019predicting} built on the aforementioned studies by using self-reported citizenship information from the 2018 ACS combined with restricted Social Security Numident administrative data to examine misreporting of citizenship and immigration status. They found that the estimated naturalized citizen population is substantially larger when naturalization is determined based on self-reported citizenship than when the naturalized population is defined using administrative records. Non-citizens misreported their own citizenship status and the status of other household members in surveys. \citet{brown2019predicting} assume the misreporting is due to misunderstanding the question or feeling an incentive to misreport to surveyors.

In sum, social scientists have long suspected that some immigration-specific survey measures, such as citizenship status, are substantially misreported. These concerns have been buffeted by a small number of studies attempting, through indirect methods, to estimate the scale of misreporting. To date, we know little to nothing about the selectivity of citizenship misreporting and, thus, whether it biases estimates of the association between citizenship and key integration outcomes such as health insurance coverage. 

In this paper we propose a Bayesian modeling approach for unit-level complex survey data that accommodates binary responses assumed to be measured with error, for the purpose of estimating the health insurance coverage gap between naturalized citizens and noncitiziens in the U.S., in the presence of misreporting citizenship status. Although the statistics literature on measurement error modeling is extensive (e.g., see \cite{carroll2006measurement},  \cite{sinha2021bayesian}, and the references therein), the intersection with survey methodology is significantly less developed by comparison. Our methodological contribution resides squarely within this intersection.

In the context of complex survey data, the measurement error methodology is typically applied to area-level (i.e., tabulation-level) data. For this type of modeling the response is usually a design-based (direct) estimate (e.g., a survey weighted estimate (Horvitz-Thompson estimate) for a total). In addition to the direct estimate, typically an estimate of the sampling error variance is provided.

An early example of measurement error modeling in the context of survey methodology is provided by \cite{ybarra2008small}. Their paper proposes a Fay-Herriot model for small area estimation, where the auxiliary information (covariates) are measured with error. Subsequent examples of measurement error modeling in the survey setting includes \citet{arima2015bayesian}, \citet{arima2017multivariate},  among others. Going a step further, \citet{nandy2022bayesian} considers multi-type survey data (e.g., Poisson, Gaussian, binomial, etc.) measured with spatially correlated covariates measured with error. Although there is a growing body of literature, there are substantially fewer examples, by comparison, of accounting for measurement error in the unit-level modeling setting.

One example of measurement error modeling at the unit-level is \citet{da2016using}. This research focuses on using binary paradata to identify the presence of measurement error and proposes two estimation approaches that address the complex survey designs -- pseudo-maximum likelihood estimation and parametric fractional imputation. More recently, \citet{da2021testing} propose a test for measurement error using paradata. Also, \cite{arima2019unit} consider unit-level small area estimation with misclassified covariates, although only for the case of an ignorable survey design. { Another approach to handling measurement error for binary complex survey data is proposed by \citet{so2020correlated}, where the authors introduce a generalized estimating equation (GEE) approach. Their approach focuses on cluster sampling and develops analysis strategies for analyzing binary survey responses. Uncertainty estimation relies on the GEE framework and is, therefore, less tractable than our approach in finite samples.} There are several other examples of unit-level models using survey weights in the presence of covariate measurement error; e.g., see \citet{torabi2011small}, \citet{biemer2010total}, and the references therein. The latter reference \citep{biemer2010total} focuses on the {\it total survey error} paradigm in which interest resides in measuring all sources of error that arise in the design, collection, processing and analysis of survey data.

The remainder of this paper proceeds as follows. Section~\ref{sec:meinsd} introduces the developed statistical model for addressing measurement error in the context of survey data. In addition, this section establishes notation and presents results of a model-based simulation study, highlighting the effectiveness of our proposed methodology. Section~\ref{sec:analysis} presents an analysis of the effect of citizenship status on health insurance and uses data from the Survey of Income and Program Participation (SIPP) collected by the U.S. Census Bureau. Importantly, this analysis is a driving factor behind our methodological development, as our goal is to evaluate the health insurance coverage gap between naturalized citizens and noncitiziens in the U.S., in the presence of misreporting citizenship status.  Finally, we provide concluding discussion in Section~\ref{sec:disc}.

\section{Measurement Error in Survey Data}\label{sec:meinsd}

We begin by considering a binary response, $Y_i$, associated with the $i$th respondent in a sample survey. Along with the response, we observe a $p$-dimensional covariate vector, $\bm{x}_i$, as well as a survey weights, $w_i$, that is assumed to be proportional to the probability of selection. One common challenge is that the binary response may be measured with error. The use of these types of error-prone variables, as either a response or covariate, can lead to bias in standard statistical models. 

Although the literature on misclassified binary data in the survey setting is sparse, this problem has seen more attention in fields such as disease testing \citep{vansteelandt2000regression, mcmahan2017bayesian} and ecology \citep{dorazio2012gibbs}.  Assuming an ignorable survey design, one natural approach to handle this problem is to use a data augmented likelihood,
\begin{align*}
    p(\bm{Y}, \bm{\widetilde{Y}} | \bm{S}_e, \bm{S}_p,\bm{p}) &= \prod_{i=1}^n \left\{\bm{S}_e^{Y_i} (1-\bm{S}_e)^{1-Y_i} \right\}^{\widetilde{Y}_i} \left\{(1-\bm{S}_p)^{Y_i}\bm{S}_p^{1-Y_i} \right\}^{1-\widetilde{Y}_i} \\
    & \times \prod_{i=1}^n p_i^{\widetilde{Y}_i} (1-p_i)^{1-\widetilde{Y}_i}, 
\end{align*} where $\widetilde{Y}_i$ is the latent true outcome for unit $i$ in the sample and $Y_i$ is the observed error-prone outcome. Here , $\bm{S}_e$ is the sensitivity, or $P(Y_i=1|\widetilde{Y}_i=1)$ and $\bm{S}_p$ is the specificity or $P(Y_i=0|\widetilde{Y}_i=0).$ Finally, $p_i$ is the probability of occurrence for unit $i$, which can be related to covariates through further modeling.

We rewrite this likelihood hierarchically as 
\begin{equation}\label{eq: ign}
    \begin{split}
        Y_i | \widetilde{Y}_i, \bm{S}_e, \bm{S}_p & \stackrel{ind}{\sim} \left\{\bm{S}_e^{Y_i} (1-\bm{S}_e)^{1-Y_i} \right\}^{\widetilde{Y}_i} \left\{(1-\bm{S}_p)^{Y_i}\bm{S}_p^{1-Y_i} \right\}^{1-\widetilde{Y}_i} \\
    \widetilde{Y}_i | \bm{p}  & \stackrel{ind}{\sim} \hbox{Bernoulli}(p_i).
    \end{split}
\end{equation} Rewriting as such allows us to interpret the observed response data as a mixture of two Bernoulli distributions, where the mixture component is determined by the true response value. Within each component, the probabilities of occurrence are defined by either the sensitivity or specificity.  

Model~\ref{eq: ign} assumes that the survey design is ignorable. However, in practice complex survey designs often result in an informative sample, where a unit's probability of selection is associated with the response. Assuming an ignorable design in these cases can result in substantial biases \citep{pfe07}. In order to mitigate this source of bias, the survey design should be accounted for in some way, typically via the survey weights. \cite{parker2023comprehensive} provide a review of the various ways to account for informative sampling; however, one common approach is to use a pseudo-likelihood \citep{bin83, ski89}. A pseudo-likelihood works by re-weighting each unit's likelihood contribution according to its survey weight,
$$
\prod_{i \in \mathcal{S}} f(Y_i|\bm{\theta})^{w_i}.
$$ The pseudo-likelihood can be maximized; however, \cite{sav16} show that the pseudo-likelihood can be used in a general Bayesian setting, after rescaling the survey weights to sum to the sample size for correct uncertainty quantification. The Bayesian pseudo-likelihood has been used to model binary/categorical survey data without error \citep{parker2022computationally}, however not in the context of error-prone or misclassified data.

Thus, to extend Model~\ref{eq: ign} to account for informative sampling, we introduce a weighted pseudo-likelihood,
\begin{align*}
    Y_i | \widetilde{Y}_i, \bm{S}_e, \bm{S}_p & \propto \left\{\bm{S}_e^{Y_i} (1-\bm{S}_e)^{1-Y_i} \right\}^{\widetilde{w}_i\widetilde{Y}_i} \left\{(1-\bm{S}_p)^{Y_i}\bm{S}_p^{1-Y_i} \right\}^{\widetilde{w}_i(1-\widetilde{Y}_i)} \\
    \widetilde{Y}_i | \bm{p}  & \propto \left\{\hbox{Bernoulli}(\widetilde{Y}_i | p_i)\right\}^{\widetilde{w}_i},
\end{align*} where $\widetilde{w}_i$ denotes the reported survey weight after scaling to sum to the sample size. Finally, after placing conjugate prior distributions on the sensitivity and specificity, as well as modeling the true responses with covariates, our full model hierarchy is
\begin{align*}
    Y_i | \widetilde{Y}_i, \bm{S}_e, \bm{S}_p & \propto \left\{\bm{S}_e^{Y_i} (1-\bm{S}_e)^{1-Y_i} \right\}^{\widetilde{w}_i\widetilde{Y}_i} \left\{(1-\bm{S}_p)^{Y_i}\bm{S}_p^{1-Y_i} \right\}^{\widetilde{w}_i(1-\widetilde{Y}_i)} \\
    \widetilde{Y}_i | \bm{p}  & \propto \left\{\hbox{Bernoulli}(\widetilde{Y}_i | p_i)\right\}^{\widetilde{w}_i} \\
    \hbox{logit}(p_i) &= \bm{x}_i'\bm{\beta} \\
    \bm{S}_e & \sim \hbox{Beta}_{0.5}(\alpha_e, \beta_e) \\
    \bm{S}_p & \sim \hbox{Beta}_{0.5}(\alpha_p, \beta_p) \\
    \bm{\beta} & \sim \hbox{N}(\bm{0}, \sigma_{\beta}^2 \bm{I}),
\end{align*} where $\hbox{Beta}_{0.5}(a, b)$ corresponds to a Beta distribution truncated below at 0.5. This is done to allow for identifiability of the error rates. In other words, we enforce that the error rates be less than 0.5, which may not be the case in every situation. Nonetheless, situations where people are more likely to misreport than report correctly are extremely difficult to handle, and are beyond the scope of this work. This model allows us to estimate the unbiased effects of the covariates simultaneously with the error rates, while also providing estimates of the true response values as a byproduct.

\subsection{Multiclass Responses}

The framework used for binary data with misclassification can be extended to handle multiclass response types (i.e. Multinomial/categorical data). For example, rather than using a mixture of Bernoulli distributions, we can use a mixture of Multinomial distributions. In this case, for a response type with $K$ categories, instead of sensitivity and specificity, we must model a set of reporting rates, $r_{k_1 k_2}$ for $k_1,k_2=1,\ldots,K,$ where $r_{k_1 k_2}$ represents the probability of class $k_1$ being reported given that the true class is $k_2.$ Then, given a true class of $k,$ the observed response can be represented as a Multinomial random variable with size one and probability vector $\bm{r}_{k} = (r_{1 k},\ldots, r_{K k})'.$

The full multiclass model can be written hierarchically as
\begin{align*}
    Y_i | \widetilde{Y}_i, \bm{R} & \propto \prod_{k=1}^K \hbox{Multinomial}(1, \bm{r}_k)^{\widetilde{w}_i I(\widetilde{Y}_i=k)}  \\
    \widetilde{Y}_i | \bm{p}_i & \propto \hbox{Multinomial}(1, \bm{p}_i)^{\widetilde{w}_i} \\
    p_{ik} & = \frac{\hbox{exp}(\bm{x}_i'\bm{\beta}_k)}{\sum_{k=1}^K \hbox{exp}(\bm{x}_i'\bm{\beta}_k)} \\
    \bm{\beta}_k & \sim \hbox{N}(\bm{0}, \sigma^2 I), \; k=1,\ldots,K-1 \\
    \bm{r}_k & \sim \hbox{Dirichlet}\left(\frac{1}{K},\ldots,\frac{1}{K}\right), \; k=1,\ldots,K
\end{align*} Note that $\bm{\beta}_K$ is set to be equal to the zero vector for identification purposes.

\subsection{Simulation}

To illustrate this approach, we conduct a short simulation study. We construct a population of 100,000 individuals, with four covariates. All covariates are generated independently from a standard normal distribution.  We generate true success probabilities as $\hbox{logit}^{-1}(\bm{x}_i'\bm{\beta})$, where $\bm{\beta}=(0.7,-2.0,0.5, -0.3)'.$ Finally, we use these probabilities to sample the true response values (i.e., 1 could correspond to a citizen and 0 could correspond to a non-citizen) from independent Bernoulli distributions, denoted as $\widetilde{Y}_i$. 

After generating a vector of true citizenship statuses, we add measurement error by randomly swapping ones with zeroes using probability $(1-\bm{S}_e)$ and randomly swapping zeroes with ones using probability $(1-\bm{S}_p).$ Here, we let one misreporting rate be larger than the other, by setting $\bm{S}_e=0.9$ and $\bm{S}_p=0.75.$ That is,  we expect that most citizens will respond as such, but a higher proportion of non-citizens will respond as citizens. We denote these error-prone response values as $Y_i.$

After constructing the population, we take an informative sample using Poisson sampling with probability proportional to $\hbox{exp}(z_i - \widetilde{Y}_i)^{0.1}$, where $z_i \stackrel{iid}{\sim} \hbox{N}(0,1),$ and an expected sample size of 1,000. The use of $z_i$ here represents ignorable sampling variation; however, the use of $\widetilde{Y}_i$ enforces an informative sample. For the sampled individuals, we observe the covariates, $\bm{x}_i$, the error-prone response, $Y_i$, and the survey weights, $w_i.$ We fit the proposed (Mixture) model as well as a model that accounts for informative sampling, but not measurement error (e.g., see \cite{parker2022computationally}, denoted as Naive). 
We repeat this entire process 100 times in order to compare  bias for the estimates of $\bm{\beta}.$ 

\begin{figure}[H]
    \begin{center}
        \includegraphics[width=150mm]{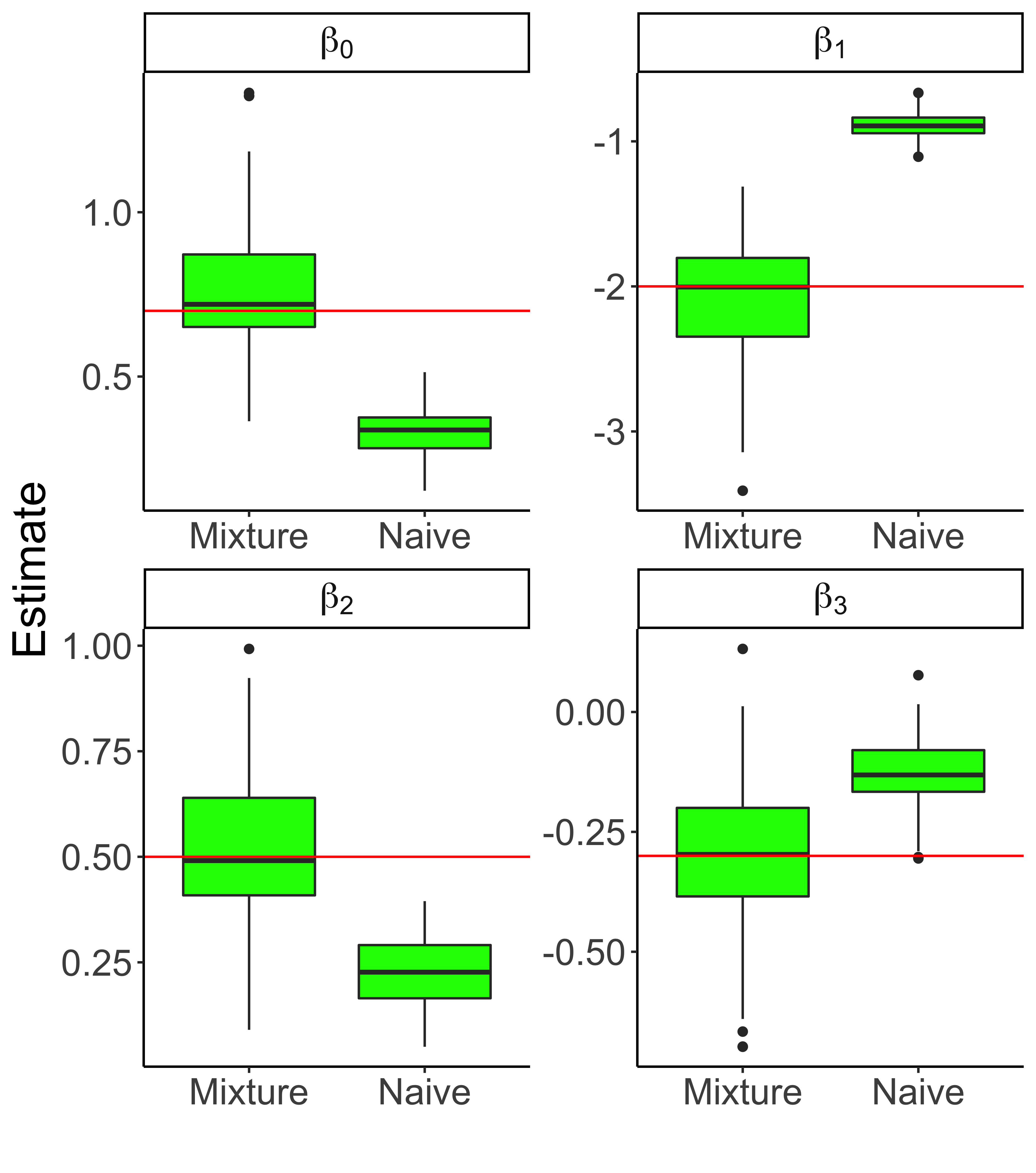}
         \caption{\baselineskip=10pt Boxplots of posterior mean regression coefficient estimates under the mixture corrected vs. naive models. True parameter values are shown as red horizontal lines.}
         \label{fig: violin}
    \end{center}               
\end{figure}

Figure~\ref{fig: violin} compares boxplots of the posterior mean for each regression coefficient across all 100 simulated datasets, along with the true parameter values. There is a clear bias for each regression coefficient when considering the naive model. However, the proposed mixture model is able effectively eliminate this bias, as evidenced by the boxplots being centered around the true data generating values. One point of note is that there is additional uncertainty induced by the mixture model, which is reflected in the larger spread for the mixture model boxplots.

\section{Analysis of Citizenship Status Effect on Health Insurance}\label{sec:analysis}

Social scientists are increasingly interested in differences in immigrants’ socioeconomic adaptation that stem from variation in rights and privileges inhering in one’s citizenship status. Legally, naturalized citizens enjoy equal rights and eligibility for public programs as any U.S. citizen. Therefore, among immigrants, the difference in material wellbeing between naturalized citizens and noncitizens is interpreted at least in part as deriving from the benefits of greater legal and civic inclusion. Health insurance is an important indicator of socioeconomic integration, both as an outcome in its own right, but also because it is strongly linked to the prevalence other health outcomes, that, in turn threaten social mobility and material wellbeing. Research has documented large differences in health insurance coverage between naturalized citizens and noncitizens, but to date, no study has considered the potential that misreporting of citizenship may bias the estimated gaps.

Thus, our goal is to study the health insurance coverage gap between naturalized citizens and noncitiziens in the U.S., while accounting for the bias attributable to misreporting of citizenship. To do so, we utilize the Survey of Income and Program Participation (SIPP), which presents the additional challenge of accounting for the survey design, in order to prevent bias due to an informative sample. 

\subsection{Data Description}

The SIPP is a national longitudinal survey conducted by the U.S. Census Bureau. The goal of the survey is to yield information related to important topics such as employment, participation in government programs, and measures of economic wellbeing (e.g., education, health insurance, food security, etc.).\footnote{\url{https://www.census.gov/programs-surveys/sipp/about.html}} The survey uses a complex two-stage sampling design, where primary sampling units (counties or groups of counties) are first selected, and then individual households are selected within the primary sampling units. Survey weights are provided, corresponding to inverse probabilities of selection, with potential adjustments for issues such as nonresponse.

In order to study the impact of citizenship status on health insurance, we use data from the 2018--2021 waves of the SIPP.  Specifically, we construct a cross-sectional dataset by aggregating the first waves of each of the 2018, 2019, 2020, and 2021 panels of the SIPP, as citizenship status is only asked during the first wave. This yields a sample that is representative of the 2018--2021 period and results in a sample size of 65,434 of which 10,492 are foreign-born individuals. We note that we scale the reported survey weights by a factor of one fourth to account for the aggregation of multiple panels \citep{lumley2011complex}. Survey respondents were asked to respond to questions on a number of different topics, including citizenship status, which is subject to misreporting error.

\subsection{Methodology}

Using data from the SIPP, we estimate the effect on health insurance attributable to citizenship status for the U.S. foreign-born population. As a measure of citizenship status, we use a binary variable that indicates whether or not an individual is a naturalized citizen (1 = yes, 0 = no). Our primary response of interest is the binary variable that indicates whether or not an individual had health insurance coverage at the time of the survey (1 = yes, 0 = no). Here, health insurance coverage includes both public and private forms of health insurance.

As a baseline, we fit a naive logistic regression with a pseudo-likelihood to account for the unequal probabilities of selection in the survey,
\begin{align*}
    \bm{y} &\propto \prod_{i \in \mathcal{S}} \hbox{Bernoulli}(y_i | p_i)^{\widetilde{w}_i}\\
    \hbox{logit}^{-1}(p_i) &= \bm{x}_i'\bm{\beta},
\end{align*} where $\hbox{Bernoulli}(y_i | p_i)$ represents a Bernoulli probability mass function with probability $p_i$ evaluated at $y_i.$ In this case, $y_i$ represents the health insurance status of respondent $i$. The vector of covariates $\bm{x}_i$ includes an intercept, categorical covariates responding to race, sex, marital status, child status, education level, veteran status, union membership, and government temporary status, as well as the integer valued covariates age and household size. These covariates are used as control variables. Finally, the vector of covariates includes the binary citizenship status variable ($x_{ci}$). Although it is known that this citizenship status variable contains substantial error, the naive model does not account for this. The regression coefficient corresponding to citizenship status ($\beta_c$) is the primary parameter of interest. 

In addition to the naive model, we fit our proposed mixture corrected model
\begin{align*}
        \bm{y}|\bm{p}_h &\propto \prod_{i \in \mathcal{S}_f} \hbox{Bernoulli}(y_i | p_{hi})^{\widetilde{w}_i} \\
    \bm{x}_{c} | \widetilde{\bm{x}}_{c}, {S}_e, {S}_p & \propto \prod_{i \in \mathcal{S}} \left\{{S}_e^{x_{ci}} (1-{S}_e)^{1-x_{ci}} \right\}^{\widetilde{w}_i\widetilde{x}_{ci}} \left\{(1-{S}_p)^{x_{ci}}{S}_p^{1-x_{ci}} \right\}^{\widetilde{w}_i(1-\widetilde{x}_{ci})} \\
    \widetilde{\bm{x}}_{c} | \bm{p}_c  & \propto  \prod_{i \in \mathcal{S}} \hbox{Bernoulli}(x_{ci} | p_{ci})^{\widetilde{w}_i} \\
    \hbox{logit}(p_{hi}) &= \bm{x}_{1i}'\bm{\beta}_1 \\
    \hbox{logit}(p_{ci}) &= \bm{x}_{2i}'\bm{\beta}_2 \\
    {S}_e & \sim \hbox{Beta}_{0.5}(\alpha_e, \beta_e) \\
    {S}_p & \sim \hbox{Beta}_{0.5}(\alpha_p, \beta_p) \\
    \bm{\beta}_1 & \sim \hbox{N}(\bm{0}, \sigma_{\beta}^2 \bm{I}) \\
    \bm{\beta}_2 & \sim \hbox{N}(\bm{0}, \sigma_{\beta}^2 \bm{I}).
\end{align*} Here, $y_i$ represents the health insurance status of respondent $i$, $x_{ci}$ represents the observed (error-prone) citizenship status, and $\widetilde{x}_{ci}$ represents the latent true citizenship status. Our primary goal is inference on the relationship between health insurance and citizenship status for the foreign-born population. Thus, the health insurance model is fit using a subset of the SIPP sample data, $\mathcal{S}_f \subset \mathcal{S}$, that only includes foreign-born individuals. In other words, the vector $\bm{y}$ contains elements $y_i$ for $i \in \mathcal{S}_f.$ Meanwhile, the measurement error model for citizenship status utilizes the entire set of sample respondents. The covariates used in the health insurance model, $\bm{x}_{1i},$ are the same as those considered in the naive model, with one exception. Rather than using the observed citizenship status $x_{ci}$, our model uses the latent true but unknown status. $\widetilde{x}_{ci}$ (which is jointly estimated). The covariate vector in the citizenship status model, $\bm{x}_{2i}$, includes information about the categorical covariates entry year, birth region, English proficiency, sex, marital status, child status, education level, and home ownership. It also includes age, household size, and income to poverty ratio. 

One important issue is that the covariates in $\bm{x}_{2i}$ are highly colinear with foreign-born status, which can cause identifiability issues with $\bm{\beta}_2.$ To alleviate this, we orthogonalize the untransformed design-matrix ($\bm{x}^*_{2}$) against foreign-born status. This is done by first creating a matrix $\bm{F}$ with an intercept column and a column containing indicator variables for foreign-born status. Then, $\bm{x}^*_{2i}$ is pre-multiplied by the projection matrix $\left(\bm{I} - \bm{F}(\bm{F'F})^{-1}\bm{F}'\right)$, resulting in $\bm{x}_{2i}$. Finally, the model specification is completed by placing independent vague $\hbox{N}(0, 10)$ priors on the elements of $\bm{\beta}_1$ and $\bm{\beta}_2,$ as well as setting $\alpha_e = \beta_e=\alpha_p=\beta_p=1.$

Both models were fit using Gibbs sampling with a Markov chain of 10,000 iterations, discarding the first 1,000 draws as burn-in. Convergence was assessed through visual inspection of trace plots, with no lack of convergence detected. The sampling details for this model, including full-conditional distributions are given in Appendix A, however we note that we use Pólya-Gamma data augmentation \citep{pol13} in order to efficiently sample from the posterior distribution without the need for rejection based sampling steps.

\subsection{Results}

The primary parameter of interest in this study is $\beta_c$, which represents the gap in health insurance between citizens and non-citizens. We use the posterior mean of this parameter as a point estimate, while also constructing a 95\% credible interval as an indication of the uncertainty around the estimate. The naive model results in a posterior mean for $\beta_c$ equal to 0.83 and a 95\% credible interval of (0.72, 0.94). A summary of all estimated regression coefficients is given in Table ~\ref{Tab1}.

After fitting the mixture corrected model the estimated posterior mean for $\beta_c$ was equal to 1.15 and a 95\% credible interval was estimated as (0.85, 1.29). A summary of all regression coefficients is given in Table ~\ref{Tab2}. These results show that the estimated effect of citizenship on health insurance is considerably larger after accounting for measurement error. Furthermore, there is little overlap between the credible interval for the naive model vs. the proposed error-corrected model. In addition to this, secondary parameters of interest are $S_e$ and $S_p$. The posterior mean of $1-S_e$ was 0.0001 and the posterior mean of $1-S_p$ was 0.0681. These correspond to the estimated probability of a true citizen reporting as a non-citizen and the probability of a true non-citizen reporting as a citizen respectively. Intuitively, these values make sense, as we would expect it to be very unlikely for a citizen to report non-citizenship status, while for a variety of reasons, there may be cause for a non-citizen to report as a citizen. Furthermore, these error rates align closely with those estimated by \cite{brown2019predicting} using data from the much larger American Community Survey, rather than the SIPP.

\begin{figure}[H]
    \begin{center}
        \includegraphics[width=150mm]{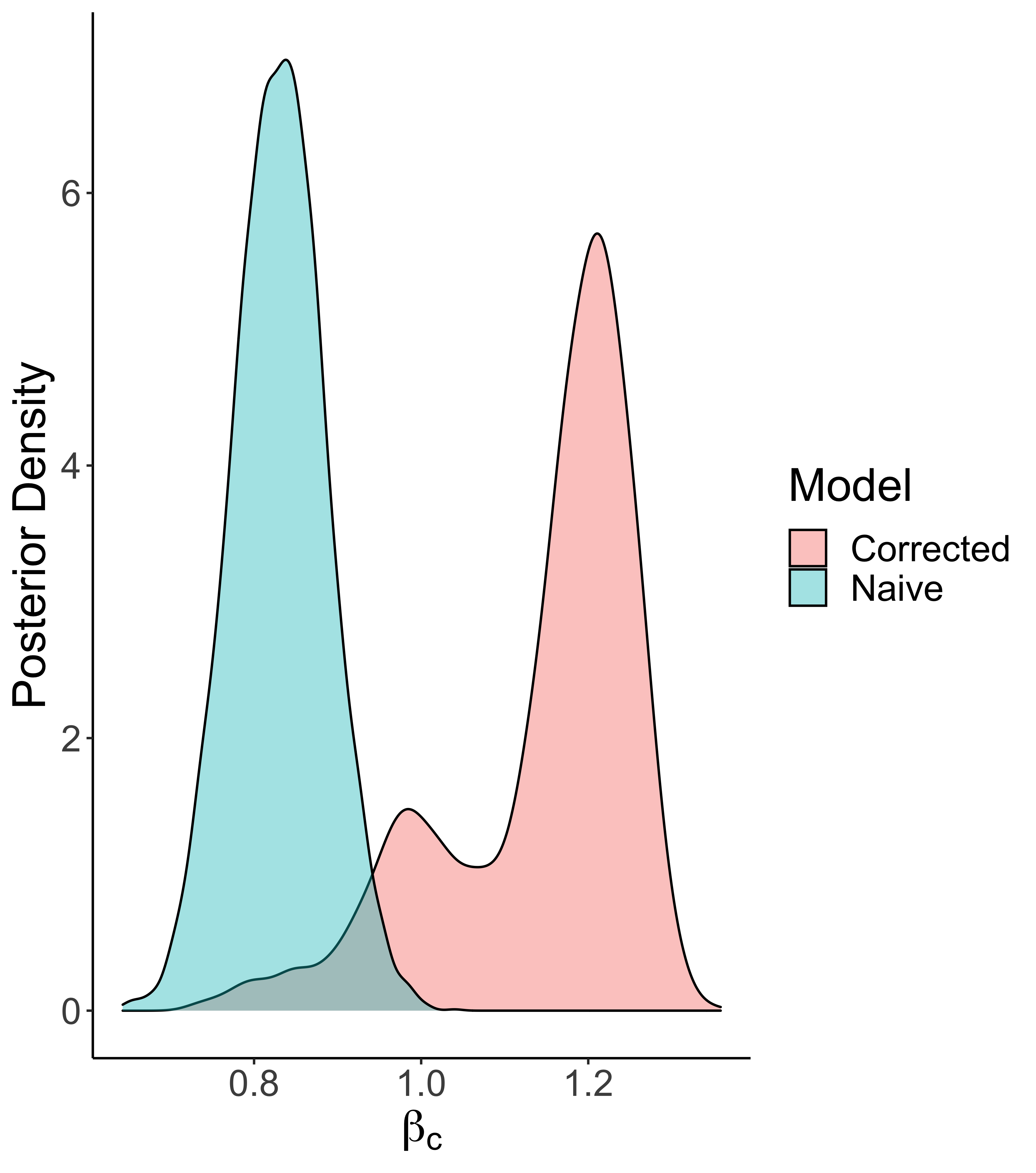}
         \caption{\baselineskip=10pt Posterior density of $\beta_c$ under the naive vs. mixture corrected model using 2018--2021 SIPP data.}
         \label{fig: coefplot}
    \end{center}               
\end{figure}

Figure~\ref{fig: coefplot} shows the marginal posterior density of $\beta_c$ under both the naive model and the corrected model. There is little overlap between the two densities, and the density under the naive model is shifted substantially to the left. This indicates that the naive approach severely underestimates the effect that citizenship has on health insurance for the foreign-born population. 

\begin{figure}[H]
    \begin{center}
        \includegraphics[width=150mm]{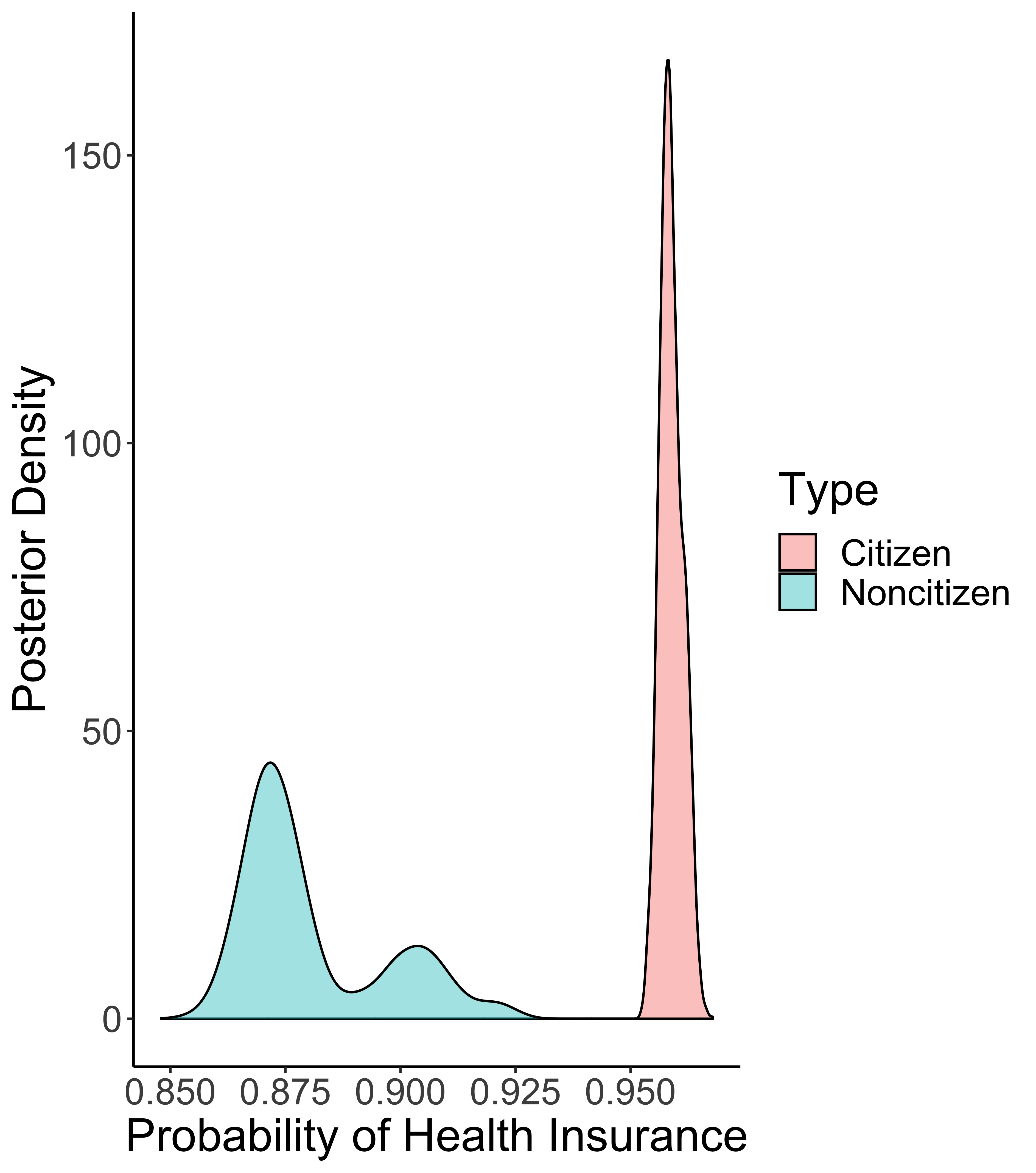}
         \caption{\baselineskip=10pt Posterior probability of health insurance for citizens vs. noncitizens based on the mixture corrected model using 2018--2021 SIPP data. All other covariates are held at their sample mode/mean. }
         \label{fig: probplot}
    \end{center}               
\end{figure}

Lastly, Figure ~\ref{fig: probplot} shows the marginal posterior density of the probability that an individual has health insurance for both citizens and noncitizens. Note that all other categorical (integer) covariates are held at their sample mode (mean) for the prediction. The noncitizen density is shifted far to the left of the citizen density, with no overlap between the two. This illustrates the importance that citizenship status plays in terms of health care coverage for the foreign-born, as noncitizens are much more likely to be without health insurance coverage.

\begin{table}

\caption{Naive model summary of estimated health insurance regression coefficients using 2018--2021 SIPP data. Results include the posterior mean, standard error (posterior standard deviation), and the 2.5 and 97.5 percentiles of the posterior distribution. The coefficient of interest ($\beta_c$) is highlighted.}
\centering
\begin{tabular}[t]{l|r|r|r|r}
\hline
Coefficient & Posterior Mean & Standard Error & 2.5\% & 97.5\%\\
\hline
(Intercept) & -0.579 & 0.155 & -0.884 & -0.272\\

race2 & 0.300 & 0.081 & 0.142 & 0.462\\

race3 & 1.030 & 0.076 & 0.883 & 1.181\\

race4 & 0.028 & 0.141 & -0.249 & 0.307\\

age & 0.016 & 0.003 & 0.011 & 0.021\\

male1 & -0.326 & 0.052 & -0.428 & -0.224\\

marst2 & -0.512 & 0.125 & -0.757 & -0.269\\

marst3 & -0.546 & 0.217 & -0.964 & -0.115\\

marst4 & -0.338 & 0.102 & -0.537 & -0.136\\

marst5 & -0.369 & 0.145 & -0.650 & -0.086\\

marst6 & -0.224 & 0.069 & -0.358 & -0.088\\

ownchild1 & 0.068 & 0.066 & -0.062 & 0.199\\

hhhsize & 0.008 & 0.018 & -0.028 & 0.043\\

educrec2 & 0.445 & 0.068 & 0.312 & 0.580\\

educrec3 & 0.725 & 0.077 & 0.575 & 0.876\\

educrec4 & 1.441 & 0.079 & 1.286 & 1.595\\

veteran1 & 0.698 & 0.340 & 0.067 & 1.397\\

union1 & 1.124 & 0.151 & 0.832 & 1.426\\

govtemp1 & 0.524 & 0.156 & 0.224 & 0.833\\

citizenship & \hl{\textbf{0.831}} & 0.056 & 0.721 & 0.941\\
\hline
\end{tabular}\label{Tab1}
\end{table}

\begin{table}

\caption{Mixture corrected model summary of estimated health insurance regression coefficients using 2018--2021 SIPP data. Results include the posterior mean, standard error (posterior standard deviation), and the 2.5 and 97.5 percentiles of the posterior distribution. The coefficient of interest ($\beta_c$) is highlighted.}
\centering
\begin{tabular}[t]{l|r|r|r|r}
\hline
Coefficient & Posterior Mean & Standard Error & 2.5\% & 97.5\%\\
\hline
(Intercept) & -0.078 & 0.123 & -0.281 & 0.205\\

race2 & -0.249 & 0.035 & -0.319 & -0.181\\

race3 & 0.517 & 0.080 & 0.340 & 0.658\\

race4 & -0.132 & 0.059 & -0.247 & -0.018\\

age & 0.013 & 0.002 & 0.011 & 0.017\\

male1 & -0.286 & 0.024 & -0.333 & -0.238\\

marst2 & -0.724 & 0.079 & -0.876 & -0.572\\

marst3 & -0.489 & 0.097 & -0.679 & -0.299\\

marst4 & -0.542 & 0.043 & -0.625 & -0.456\\

marst5 & -0.673 & 0.075 & -0.820 & -0.523\\

marst6 & -0.344 & 0.035 & -0.412 & -0.274\\

ownchild1 & 0.296 & 0.032 & 0.233 & 0.360\\

hhhsize & -0.031 & 0.009 & -0.049 & -0.013\\

educrec2 & 0.368 & 0.041 & 0.291 & 0.452\\

educrec3 & 0.730 & 0.043 & 0.648 & 0.817\\

educrec4 & 1.334 & 0.043 & 1.250 & 1.418\\

veteran1 & 0.611 & 0.074 & 0.468 & 0.759\\

union1 & 0.792 & 0.066 & 0.664 & 0.924\\

govtemp1 & 0.632 & 0.060 & 0.516 & 0.750\\

citizenship & \hl{\textbf{1.146}} & 0.116 & 0.848 & 1.287\\
\hline
\end{tabular}\label{Tab2}
\end{table}

\section{Discussion}\label{sec:disc}

Motivated by the goal of estimating the health
insurance coverage gap between naturalized citizens and noncitizens in the U.S., we develop statistical methodology to account for the measurement error that results from misreporting citizenship. Unlike the majority of literature on misclassified binary data, our approach uses a Bayesian pseudo-likelihood and, therefore, also addresses potential biases that can manifest as a result of complex survey design. In addition, we provide an extension from the binary categorical case to the case of multiclass responses.

To study the effect of citizenship status on health insurance, we utilize SIPP data from 2018--2021. Importantly, we demonstrate via simulation and through our analysis that principled accounting for misreporting of citizenship status is critical and that failure to do so could result in erroneous inference. Further, our proposed methodology fills a significant gap in the applied social science and demography literature and will assist subject matter experts in making correct inference in the case of misreporting for complex survey data.

Our results may also have signifanct policy implications. For example, we have shown that use of a naive modeling approach that does not account for measurement error results in underestimation of the health insurance gap between citizens and noncitizens. This could in turn lead to misguided policies or interventions directed at increasing insurance coverage, particularly for vulnerable groups. Given that health insurance coverage is strongly linked to health care access, utilization, and more favorable health outcomes, such misguided interventions could have negative downstream consequences such as untreated or undiagnosed health conditions. 

Finally, our results (Section~\ref{sec:analysis}) corroborate previous analyses in the demography literature \citep{brown2019predicting} that use data from the much larger American Community Survey. In addition, our analysis illustrates the importance that citizenship status plays in terms of health care coverage for the foreign-born, as noncitizens are much more likely to be without health insurance coverage. Ultimately, we expect that the approach proposed here will be widely used by subject-matter experts to analyze data from the SIPP and other complex surveys.

\section*{Acknowledgments}
  This article is released to inform interested parties of ongoing research and to encourage discussion. The views expressed on statistical issues are those of the authors and not those of the NSF or U.S. Census Bureau. This research was partially supported by the U.S. National Science Foundation (NSF) under NSF grants NCSE-2215168 and NCSE-2215169.

\newpage
\bibliographystyle{jasa}

\section*{Appendix A: Posterior Sampling Steps}
In order to easily sample from the posterior distribution of the model specified in Section~\ref{sec:analysis} we use data augmentation. Specifically, \cite{pol13} show how the introduction of Pólya-Gamma random variables can lead to conditionally conjugate structure in logistic models. A random variable $X$ is said to have a P\'{o}lya-Gamma distribution with parameters $b>0$ and $c \in \mathcal{R}$, denoted $\hbox{PG}(b,c)$, if $X$ is equal in distribution to
$$ \frac{1}{2\pi^2} \sum_{k=1}^{\infty} \frac{g_k}{(k-1/2)^2 + c^2/(4\pi^2)},$$ where $g_k \stackrel{ind}{\sim} Gamma(b,1)$. Importantly, \cite{pol13} show the identity 
\begin{equation}\label{eq: pg}
   \frac{(e^{\psi})^a}{(1 + e^{\psi})^b} = 2^{-b}e^{\kappa \psi} \int_0^{\infty} e^{-\omega \psi^2/2} p(\omega) d\omega, 
\end{equation} where $\kappa = a - b/2$ and $p(\omega)$ is a $\hbox{PG}(b,0)$ density. Thus, conditional on the latent variable $\omega,$ the logistic likelihood is proportional to a Gaussian density. We note that our models for $\bm{y}$ and $\bm{\widetilde{x}}_c$ are both logistic models. Thus, through the introduction of two different sets of latent PG random variables, we can attain conditional conjugacy via our Gaussian prior distributions for $\bm{\beta}_1$ and $\bm{\beta}_2$.

We introduce the latent variables $\omega_{1i}$ for $i \in \mathcal{S}_f$ and $\omega_{2i}$ for $i \in \mathcal{S}.$ Suppose the there are $n_1$ samples in $\mathcal{S}_f$ and $n_2$ samples in $\mathcal{S}$. Let $\bm{\omega}_1=(\omega_{11}, \ldots, \omega_{1n_1})'$ and $\bm{\omega}_2=(\omega_{21}, \ldots, \omega_{2n_1})'$, while $\bm{\Omega}_1=\hbox{Diag}(\bm{\omega}_1)$ and $\bm{\Omega}_2=\hbox{Diag}(\bm{\omega}_2)$. Furthermore, let $\bm{\kappa}_1=\left(\widetilde{w}_{1}(y_1 - 0.5), \ldots, \widetilde{w}_{n_1}(y_{n_1} - 0.5)\right)'$  and $\bm{\kappa}_2=\left(\widetilde{w}_{1}(\widetilde{x}_{c1} - 0.5), \ldots, \widetilde{w}_{n_1}(\widetilde{x}_{c n_2} - 0.5)\right)'$. 
Finally we construct the $n_1 \times P_1$ design matrix $\bm{X}_1,$ where row $i$ is equal to the vector $\bm{x}_{1i}$ along with the 
$n_2 \times P_2$ design matrix $\bm{X}_2,$ where row $i$ is equal to the vector $\bm{x}_{2i}$.

In order to sample from the posterior distribution of our model parameters given the data, we use Gibbs sampling by iteratively sampling from the below full-conditional distributions. Note that each of these distributions is of a known family that is straightforward to sample from.
\begin{enumerate}
    \item $\omega_{1i} | \cdot  \sim PG(\widetilde{w}_i, \bm{x}_{1i}'\bm{\beta}_1), \; i \in \mathcal{S}_f$
    \item $\omega_{2i} | \cdot  \sim PG(\widetilde{w}_i, \bm{x}_{2i}'\bm{\beta}_2), \; i \in \mathcal{S}$
    \item $S_p | \cdot \sim \hbox{Beta}_{0.5}(\alpha_p + \sum_{i \in \mathcal{S}}(1-x_{ci})(1-\widetilde{x}_{ci})\widetilde{w}_i, \; \beta_p + x_{ci}(1-\widetilde{x}_{ci})\widetilde{w}_i)$
    \item $S_e | \cdot \sim \hbox{Beta}_{0.5}(\alpha_e + \sum_{i \in \mathcal{S}}x_{ci}\widetilde{x}_{ci}\widetilde{w}_i, \; \beta_p + (1-x_{ci})\widetilde{x}_{ci}\widetilde{w}_i)$
    \item $\bm{\beta}_1 | \cdot \sim \hbox{N}\left(\bm{\mu} = \left(\bm{X}_1'\bm{\Omega}_1\bm{X}_1 + \frac{1}{\sigma^2_{\beta}} \bm{I} \right)^{-1} \bm{X}_1'\bm{\Omega}_1 (\bm{\kappa}_1 / \bm{\omega}_1), \bm{\Sigma}= \left(\bm{X}_1'\bm{\Omega}_1\bm{X}_1 + \frac{1}{\sigma^2_{\beta}} \bm{I} \right)^{-1} \right)$
    \item $\bm{\beta}_2 | \cdot \sim \hbox{N}\left(\bm{\mu} = \left(\bm{X}_2'\bm{\Omega}_2\bm{X}_2 + \frac{1}{\sigma^2_{\beta}} \bm{I} \right)^{-1} \bm{X}_2'\bm{\Omega}_2 (\bm{\kappa}_2 / \bm{\omega}_2), \bm{\Sigma}= \left(\bm{X}_2'\bm{\Omega}_2\bm{X}_2 + \frac{1}{\sigma^2_{\beta}} \bm{I} \right)^{-1} \right)$
        \item $\widetilde{x}_{ci} | \cdot \hbox{Binomial}\left(1, \frac{\hbox{logit}^{-1}\left(\bm{x}_{2i}'\bm{\beta}_2\right)^{\widetilde{w}_i}\bm{S}_e^{\widetilde{w}_i x_{ci}}(1-\bm{S}_e)^{\widetilde{w}_i (1-x_{ci})}}{\hbox{logit}^{-1}\left(\bm{x}_{2i}'\bm{\beta}_2\right)^{\widetilde{w}_i}\bm{S}_e^{\widetilde{w}_i x_{ci}}(1-\bm{S}_e)^{\widetilde{w}_i (1-x_{ci})} + \left( 1 -\hbox{logit}^{-1}\left(\bm{x}_{2i}'\bm{\beta}_2\right)\right)^{\widetilde{w}_i}(1-\bm{S}_p)^{\widetilde{w}_i x_{ci}}\bm{S}_p^{\widetilde{w}_i (1-x_{ci})}} \right), \\ i \in \mathcal{S}$
\end{enumerate}

\end{document}